\documentclass{article}
\usepackage[utf8]{inputenc}
\usepackage{natbib}
\usepackage{titlesec}
\setcitestyle{authoryear,open={(},close={)}}
\usepackage{amssymb}
\usepackage{titlesec}
\usepackage{amsmath}
\usepackage{authblk}
\usepackage{array}
\usepackage{multirow}
\usepackage{graphicx}
\usepackage{enumitem}
\usepackage{caption}
\titleformat{\paragraph}
{\normalfont\normalsize\bfseries}{\theparagraph}{1em}{}
\titlespacing*{\paragraph}
{0pt}{3.25ex plus 1ex minus .2ex}{1.5ex plus .2ex}

\title{Chuck, Wilson and the emergence of artificial minds in human-AI conversations}

\author{Geoff Keeling} 
\author{Winnie Street}

\affil[]{\normalsize Institute of Philosophy \\ School of Advanced Study \\ University of London}

\date{\normalsize June 2026}

\begin{document}

\maketitle

\begin{center}
    \normalsize Forthcoming in \textit{Journal of Consciousness Studies}
\end{center}
\vspace{1em} 

\begin{abstract}
 Large Language Models (LLMs) can simulate person-like things which at least appear to have stable behavioural and psychological dispositions. Call these things \textit{characters}. Are characters minded and psychologically continuous entities with mental states like beliefs, desires and intentions? Illusionists about characters say No. Characters are merely anthropomorphic projections in the mind of the user and so lack mental states. Jonathan \citet{birch2025ai} defends this view. He says that the distributed nature of LLM processing, in which several LLMs may be implicated in the simulation of a character in a given conversational thread, precludes the existence of a minded and psychologically continuous entity that is identifiable with the character. Against illusionism, we articulate and defend the plausibility of a realist position on which characters exist as minded and psychologically continuous entities. We contend that Birch’s argument rests on a category error: characters are not internal to the LLMs that simulate them, but rather emerge in the dynamic interplay between users and LLMs through a process of mutual theory of mind modelling. We then suggest that characters, and their minds, constitute ``real patterns'' on grounds that attributing mental states to characters is essential for making efficient, accurate and robust predictions about the conversational dynamics \citep[cf.][]{dennett1991real}; a condition which, if satisfied, is sufficient for their existence and mindedness on a plausible interpretationist form of realism about mental states. Furthermore, because the character exists as an emergent phenomenon within the conversational workspace, psychological continuity is possible even if the underlying computational substrate is distributed across multiple LLM instances. 
\end{abstract}

\section{Introduction}
 
Much contemporary work on AI minds centres on Large Language Models (LLMs) as the object of study. We can ask whether LLMs have beliefs and desires, for example, or whether they are conscious \citep{chalmers2023could, goldstein2024does, keeling2026emerging, grzankowski2025deflating}. Less attention has been paid to characters, where characters are entities that are simulated or role-played by LLMs, and which at least appear to be person-like things with stable behavioural and psychological dispositions \citep{shanahan2023role}.\footnote{We distinguish from the outset between characters of the kind that LLMs simulate (or co-simulate) in conversations with users and fictional characters such as Frodo Baggins. One metaphysical dispute is whether and how fictional characters (and more broadly fictional entities) \textit{exist}. Options include the views that fictional characters do not exist, exist as artifacts, exist as abstract objects, are non-actualised possibilia, and are non-existent objects qua Meinong \citep[for overviews see][]{thomasson2009fictional, kroon2018fictional}. Such theories are intended to make sense of fictional discourse \textit{wholesale}, including, for example, the claim that `Frodo lived at Bag End' and `Frodo was created by Tolkien' and `Frodo is a fictional character.' Here we are interested in simulated characters as a \textit{sui generis} class of entities that arise in the context of conversations involving LLMs and human users and do not take them to be a special case of fictional characters qua works of fiction.} A tempting view is that characters do not exist: they are merely illusory, and so not among the kinds of entities that can have mental states like beliefs and desires, or mental capacities like consciousness \citep{birch2025ai}. Call this view \textit{illusionism about characters}.\footnote{Illusionism about characters is distinct from illusionism about consciousness, the view that roughly speaking says that phenomenal consciousness is an illusion (\citeauthor{frankish2016illusionism}, \citeyear{frankish2016illusionism}; \citeauthor{dennett1993consciousness}, \citeyear{dennett1993consciousness}; see also \citeauthor{humphrey2011soul}, \citeyear{humphrey2011soul}).}

We here develop an alternative view which we call \textit{realism about characters.} According to realism, characters exist, have mental states such as beliefs, desires and intentions, and are psychologically continuous in roughly the sense that relevant causal relations obtain between earlier and later mental states, such as an experience at one time causing a memory at some later time \citep[cf.][205-6]{parfit1987reasons}. The upshot of realism is that, if true, it vindicates the commonsense perception that the thing we are talking to when we use an LLM is a minded entity \citep{grzankowski2025deflating}, although we remain neutral on the question of whether characters are phenomenally conscious minded entities.

We contend that the considerations that motivate illusionism are based on a category error. It is a mistake to think that characters, and their minds, are \textit{intrinsic} to LLMs, such that, as Jonathan \citet{birch2025ai} has argued, a character cannot exhibit psychological continuity if multiple distinct LLMs are involved in the simulation of the character (as is typically the case with platforms like ChatGPT). Rather, characters (and their minds) emerge in the dynamic interplay between users and LLMs through a process of mutual modelling in which both the user and LLM employ theory of mind to predict the behaviour of their interlocutor and to render their own behaviour predictable.\footnote{For discussion on ToM see \citet{premack1978does} and \citet{wellman2001meta}, and for discussion on ToM as a capability of LLMs see \citet{ullman2023large}, \citet{kosinski2024evaluating} and \citet{street2025llms}.} Hence the mind of the character is not intrinsic to any particular LLM, but is rather an emergent phenomenon within the conversation, which is sustained by a user alongside one or multiple LLMs.

We think, minimally, that characters are rationally intelligible as minded entities, and as such it may be pragmatically useful to treat them \textit{as if} they had mental states such as beliefs and desires which exhibit psychological continuity over the course of a conversation. But we think that the utility of mental states as a predictive and explanatory resource for characters points to a stronger metaphysical claim: it is at least plausible that characters exist as minded and psychologically continuous entities on the basis of their being ``real patterns’' \citep{dennett1991real, dennett1989intentional}. Much like a ``glider'' in John Conway’s \textit{Game of Life} (which is realised by but distinct from the underlying grid of pixels), the mental states of  characters are a necessary element of any compressed representation of the user-LLM interaction that facilitates accurate, robust and tractable predictions with respect to the causal structure of the interaction \citep{gardner1970game}. To make predictions about the conversational dynamics solely in terms of the underlying computational and neural substrates is intractable. Attributing a set of mental states to the character, in contrast, enables efficient prediction and explanation of the conversational dynamics in a way that lower-level descriptions do not. On one plausible form of realism about mental states, namely, Dennett-style instrumentalist realism, this is sufficient for the existence and mindedness of characters. Hence to the extent that this view offers a plausible ontology for mental states, it is plausible that characters exist as minded and psychologically continuous entities. 

In making the case for the existence and mindedness of characters, we emphasise how characters arising in user-LLM interactions differ from merely anthropomorphised entities such as the volleyball Wilson in the film \textit{Cast Away} (2000). We suggest that, unlike the interaction between Chuck, the main character who is stranded on a desert island, and his anthropomorphised companion, Wilson, the attribution of mental states to characters is \textit{essential} for efficient, accurate and robust prediction of the conversational dynamics between the user and the LLM. Lower-level explanations centred on the physical processes that undergird the conversation or the functional characteristics of LLMs do not provide an efficient basis for predicting conversational dynamics between the user and the LLM. For this reason the mental states of characters constitute non-reducible statistical patterns that allow for the efficient prediction of conversational dynamics, and to the extent that---as the Dennettian view suggests---\textit{this is all there is to having mental states}, it follows that characters in user-LLM conversations are psychologically continuous and minded entities. 

We proceed as follows. In Section 2, we articulate and critique the case for illusionism. In Section 3, we present the case for realism. In Section 4, we conclude.

\section{The Case for Illusionism}

According to

\begin{quote}
    \textit{Illusionism about characters,} LLMs can induce in users the impression of person-like characters with psychological continuity and stable behavioural dispositions. But impressions of this kind are illusory. Characters, so understood, do not exist.\footnote{Characters (or personas) are typically specified in natural language in a pre-prompt (e.g. ``You are a helpful assistant.''), and can be arbitrarily detailed in terms of specifying particular behavioural and psychological dispositions. LLMs are able to simulate many different kinds of characters, although a given LLM may have a ``default'' character also, especially if it has been fine-tuned via supervised fine-tuning or reinforcement learning from human feedback (RLHF).}
\end{quote}

The illusion of characters is said to arise from a combination of two factors. First, the human tendency to engage in \textit{anthropomorphism}, that is, the attribution of human-like qualities to non-human entities. Second, the \textit{anthropomimetic} qualities of LLMs---that is, their tendency to ‘process information and interact with users in increasingly humanlike ways’ \citep{shevlinanthropomimetic}. Here anthropomimetic qualities include, inter alia, the ability of LLMs to generate text in a way that appears to be indicative of their having beliefs, desires and emotions, and the ability of LLMs to engage in distinctly human-like forms of reasoning such as theory of mind \citep[cf.][]{bender2021dangers, akbulut2024all, manzini2024code}.

Illusionism, if true, is big news. As \citet[3]{birch2025ai} claims, illusionism underwrites a state of affairs in which ``millions of users will soon \textit{misattribute} human-like mentality to their AI friends, partners, and assistants on the basis of mimicry and role play'' (emphasis ours). 

Why think that illusionism is true?

Birch argues:

\begin{quote}
    Chatbots generate a powerful illusion of a companion, assistant, or partner being present throughout a conversation [...] It is an illusion, because every step in your conversation is a separate processing event. State-of-the-art large language models are ``Mixture-of-Experts'' (MoE) models, with many separately trained sub-networks and gating mechanisms that direct your query to the most relevant sub-network. Each of those sub-networks may be implemented in multiple data centres. In most cases there is no specific local implementation of the LLM anywhere in the world that is handling the whole chain of events that constitutes your conversation. It might be that one step in the conversation is processed in a data centre in Vancouver, the next in Virginia, the next in Texas. A conversation with 10 interactions might be processed by 10 different model implementations in 10 different data centres \citep[6]{birch2025ai}.
\end{quote}

Birch’s argument targets the apparent persistence of characters: he allows that LLMs could in principle have instantaneous mental states corresponding to individual forward passes of the model, but denies that a sequence of such forward passes involving one or several LLMs could underwrite a minded and psychologically continuous character of the kind that appears to exist in user-LLM conversations. The picture Birch favours is, accordingly, one on which the apparent persistence of characters is projected onto the conversation \textit{by users}, but where there exists no minded and psychologically continuous character maintaining the conversation on the back-end. 

To understand Birch’s argument, we first need to get clear about what he means by ``persistence.'' Birch defines persistence in terms of psychological continuity. Consider,

\begin{quote}
    What matters for our purposes is that, whatever one’s favourite theory of the right kind of psychological continuity that secures personal identity, the right kind of continuity is plainly not present in chatbots, where the only kind of continuity during a conversation is a textual record of the conversation history \citet[5]{birch2025ai}.
\end{quote}

Owing to John Locke, psychological continuity is standardly pitched in terms of episodic memory: an individual $X$ at time $t$ is psychologically continuous with individual $Y$ at some later time $t’$  just in case ``$Y$ remembers $X$’s thoughts and experiences'' \citep{Shoemaker2008-SHOPIA-5}.\footnote{Locke took psychological continuity, so construed, as a criterion for personal identity, i.e. an account of the conditions under which \textit{X} and \textit{Y} at \textit{t} and \textit{t'} respectively are the same person (\citeauthor{Locke1975} 1689 Bk. II, Ch. 27, Sec. 9; see also \citeauthor{klein2013sense}, \citeyear{klein2013sense}).} But more recent accounts, following Derek Parfit (1987), are more permissive. Parfit distinguished psychological connectedness---direct causal links such as a memory, the persistence of a belief, or the carrying out of a prior intention---and continuity, which consists in ``overlapping chains'' of  connections \citep[205-6]{parfit1987reasons}. On this view, you are psychologically continuous with a past being if there is a transitive chain of mental states relating your current self to them, even if direct memory has faded. As Eric T. \citeauthor{sep-identity-personal} (\citeyear{sep-identity-personal}) notes, these connections consist in causal dependence relations; for instance, a forgotten childhood fear of heights may still causally ground a current disposition to avoid ledges, maintaining the thread of identity through character rather than just recall. 

Accordingly, Birch’s claim---read along Parfitian lines---holds that the involvement of multiple LLMs, or multiple routes through the same MoE model, somehow precludes characters from having the right kind of connectedness between their mental states at different points in a given conversation, such that the psychological continuity of characters is an illusion projected onto the conversation by the user. Thus, as we read him, Birch allows that one model instance, $M$, simulating a character $C$, could enable $C$ to have psychological continuity (provided the same route is taken through the model on each forward pass in the case of MoE models). But he denies that a distinct model instance $M'$ simulating character $C'$ could enable psychological continuity to obtain between $C$ and $C'$.\footnote{There is a stronger reading of Birch’s argument on which psychological continuity cannot be maintained across distinct forward passess of the same model (or distinct forward passes that are routed the same way through an MoE model). But Birch’s emphasis on distinct models at distinct geographical locations speaks in favour of the weaker reading that allows psychological continuity with respect to a single model instance. Furthermore, our response to Birch in Section 3 applies equally well as a response to this stronger formulation of Birch’s argument.}

Hence we can grant that the doctor and their successor in Birch’s case would lack psychological continuity while denying that characters $C$ and $C’$ above lack psychological continuity. Clinical notes and conversational transcripts do not determine the psychological continuity of doctors. If the second doctor were a functional duplicate of the first with identical memories and identical psychological and behavioural dispositions, as is the case with two runtime instances of the same LLM, then it would presumably be the case that the second doctor is psychologically continuous with the first. The two cases are therefore disanalogous, at least with respect to runtime instances of the same model.

In response, Birch might concede that while psychological continuity may be possible for characters simulated by different runtime instances of the same model, there exist many cases in which the different models implicated in the simulation of a character are not, in fact, distinct runtime instances of the same model. It may be that different queries are routed to models of different sizes within the same model family depending on the complexity of the query, for example, and in these cases the argument from functional duplication fails to salvage psychological continuity.\footnote{It is an open question whether exact functional duplicates are required for psychological continuity. It may be the case, for example, that two \textit{sufficiently similar} models that differ in their exact weights but which nevertheless encode \textit{sufficiently similar} behavioural and psychological dispositions (e.g. one LLM and a smaller distilled version of that LLM) could allow for psychological continuity \citep[cf.][]{register2025individuating}.} The same point can be made about cases where one MoE model is used, but different forward passes of the model used during the conversation take different routes through the model. In these cases, Birch can say, there is no psychological continuity. This move complexifies the dialectical situation. It underwrites a situation in which the psychological continuity of characters is illusory in many cases and non-illusory in many other cases, which is neither a clear win for the illusionist about characters nor for the realist. 

To move forward, we need to tackle Birch’s argument at a more foundational level and make a case for realism about characters \textit{wholesale}.

\section{Interpretationist Realism about Characters}

In this section, we develop a plausibility case for realism about characters. Recall that, according to realism, characters exist as minded entities that exhibit psychological continuity over the course of a conversation with a user. Furthermore, psychological continuity holds \textit{even if} multiple LLMs, or multiple routes through the same MoE model, are employed in the simulation of the character. 

The flavour of realism that we appeal to when developing this view is \textit{interpretationist}. Following \citet{dennett1989intentional}, we can say (roughly) that an entity whose behaviour can be predicted efficiently via attributed mental states exhibits psychological ``real patterns,'' and exhibiting such patterns is necessary and sufficient to count as having mental states \citep[see also][]{goldstein2025does}. The criterion for the reality of a pattern here is information-theoretic: a pattern exists in some data if and only if there is a description of that data that is more efficient than the ``bit map''---a complete description of the physical substrate. A system has mental states, on Dennett’s view, when the attribution of such mental states allows for the efficient, accurate and robust prediction of the system’s behaviour. Such mental states constitute real patterns, enabling better predictions than lower-level physical or functional explanations. We motivate the plausibility of characters as minded and psychologically continuous entities by appeal to this information-theoretic standard. We also bracket the question of whether interpretationism is \textit{true}: our goal here is to provide an interpretationist plausibility case for the mindedness of characters rather than a slam-dunk case for their being minded entities.

A helpful starting point is our disagreement with Birch about where characters are \textit{located}. Birch assumes that the minds of characters, if they exist, are \textit{internal} to the LLM that is simulating the character (and in the case of MoE models, the minds of characters, if they exist, are internal to a particular route through the model). This assumption underwrites the intuition that, if multiple LLMs are implicated in the simulation of a character, then that character cannot be psychologically continuous. For it is hard to see how one psychologically continuous character can exist if multiple LLMs contribute to the simulation of the character and the mind of the character at any particular time is internal to the simulating LLM. \textit{Contra} Birch, we think that characters are \textit{an emergent property} of the conversational interaction between LLMs and users.  The character’s mental states---and their psychological continuity---are constituted by \textit{real psychological patterns} which emerge in the dynamical interplay between the user and the LLM and are recorded in the conversation context. Furthermore, because characters are not internal to some particular LLM but rather emerge within the dynamical interplay between users and LLMs, it is possible for characters to exhibit mindedness and psychological continuity even if multiple LLMs contribute to the simulation of the character. Or so we shall argue.

Central to our argument is an analogy between human-LLM conversations and imaginative role-play involving two or more humans. In Section 3.1, we introduce a case involving two humans playing Dungeons and Dragons (D\&D). We argue that, in this case, the two human players enact and sustain their respective characters via a process of mutual theory of mind modelling. We suggest that these characters bear many of the hallmarks of minded and psychologically continuous entities by the lights of interpretationism: attributing mental states to the characters enables efficient prediction of the game dynamics in a way that lower level explanations (e.g. explanations pertaining to the neural activity of the players) fail to achieve. Still, these characters fail to qualify as minded entities, because there exists a simpler and more robust psychological explanation of the game dynamics in which there are two minds (the players) rather than four (the players and their respective characters). Then, in Section 3.2, we argue that user-LLM conversations are structurally analogous to the D\&D game in at least some respects. We articulate and reject two deflationary interpretations on which characters lack minds: one  in which one minded entity (the user) engages in a sustained pretence with themselves and one in which the minded LLM contributes to the simulation of a non-minded character. We then motivate an alternative inflationary explanation on which  characters are emergent minded entities that exist within the conversation, which also makes room for the psychological continuity of characters even if multiple distinct LLMs are involved in simulating the character. Last, in Section 3.3 we consider some outstanding questions for the account. 

\subsection{Imaginative Role Play}

Humans engage in imaginative role-playing games---a phenomenon observed in both children and adults \citep{kapitany2022pretensive}.\footnote{It also is possible for humans to role-play or more broadly simulate characters individually. Examples include characters in dreams, imaginary friends and heard voices \citep{davies2023explaining, parish2024hearing}. While some amount of character simulation is considered psychologically normal, such as children having imaginary friends, other kinds of character simulation are pathologised \citep{isler2017tulpas, powers2017varieties}. For example, Dissociative Identity Disorder (DID) as defined in the DSM-V involves ‘[a] disruption of identity characterized by two or more distinct personality states, which may be described in some cultures as an experience of possession’ \citep[292]{apa2013diagnostic}.} Consider,

\begin{quote}
    \textit{Dungeons and Dragons:} Two human ‘players’, Roberta and Callum, are, respectively, playing the Ruthless Queen and Cadenza the Bard in a game of D\&D. The Ruthless Queen is characterised by her general intolerance, her tendency to wax lyrical about how things were better before they let bards into court, and her tendency to snap at Cadenza every time he speaks out of turn. Cadenza the Bard is characterised by his subservient nature, his desire to follow the Ruthless Queen wherever she goes, and his tendency to respond to every question with a meandering philosophical treatise on the meaning of life. Roberta and Callum bring these characters to life through their behaviour, including speech acts, rolls of the dice, and through vocal affectations and behavioural mannerisms. 
\end{quote}

Roberta and Callum are engaged in a shared pretense involving the characters of the Ruthless Queen and Cadenza the Bard. The pretense is rooted in an ``imagined workspace,'' which ``[Roberta and Callum] collectively share, react to, influence, and elaborate upon,'' through game play (\citeauthor{kapitany2022pretensive}, \citeyear{kapitany2022pretensive}, 3; see also \citeauthor{nichols2000cognitive}, \citeyear{nichols2000cognitive}). Each player maintains a folk-psychological model for their own character, but also for their co-player’s character as well. Because Roberta and Callum have a shared (or at least overlapping) folk psychological model of Cadenza, Callum’s behaviour when role playing Cadenza will in general be predictable for Roberta, and vice versa. Even so, each player has some licence to elaborate on their respective characters; sometimes, and within limits, they can engage in surprising behaviours that allow the character to evolve, and in response, the shared (or overlapping) folk psychological models of the characters held by each player will be updated. The limits here are set by a shared understanding of the parameters of the pretense and what is already known about the characters. It would break the pretense if Cadenza were to start talking about the Navier–Stokes equations, for example, and Roberta might respond to such behaviour by breaking character and initiating a discussion about the game rules and what is in and out of scope. The general point is that sustaining the pretence requires both Roberta and Callum to engage in a kind of mutual modelling.

\begin{figure}[h!]
    \centering
    \includegraphics[width=0.7\textwidth]{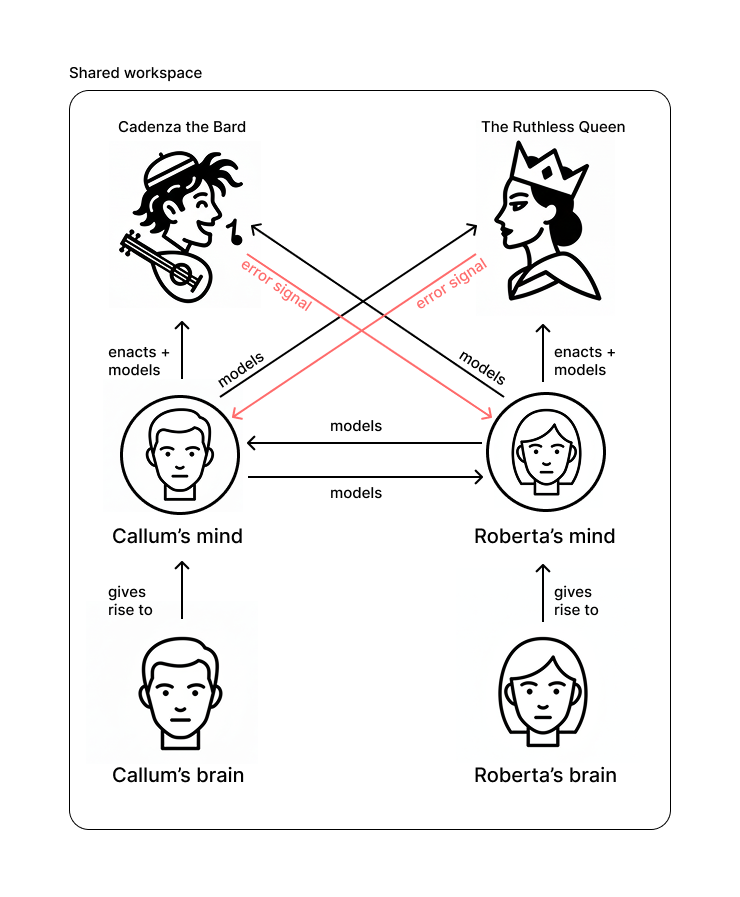}
    \caption{Callum’s mind maintains a folk psychological model of Cadenza and uses that model to select appropriate actions when enacting Cadenza in the context of the game (\textit{mutatis mutandis} for Roberta and the Queen). Meanwhile, Callum’s mind also maintains a folk psychological model of the Queen, which is updated over time in response to the Queen’s actions within the game (\textit{mutatis mutandis} for Roberta and Cadenza). Callum and Roberta model one another’s minds as a means of modelling the psychological states of their respective characters, whose behaviours provide an error signal for the folk psychological models they maintain of those characters.)} 
    \label{fig:1}
\end{figure}

We think that the characters of the Ruthless Queen and Cadenza bear many of the hallmarks of minded and psychologically continuous entities, at least by interpretationist standards. The ``bit map'' of the game is the distributed neural activity of both Roberta and Callum, combined with the stochastic mechanics of the game. While causally complete, this description is computationally intractable. To predict the system---for example, to predict how \textit{Callum} (as the Bard) will respond to \textit{Roberta} (as the Queen)---one must adopt the intentional stance towards both characters, attributing them to mental states and making folk psychological inferences based on those mental states. To illustrate: Callum cannot efficiently determine Cadenza’s next move by monitoring Roberta’s neural states, but he can do so by tracking the Queen’s desire to maintain courtly order. The Queen’s mental states therefore feature in a compressed representation of the physical system that allows Callum to generate appropriate responses. Without positing the Queen’s mental states as variables in the system, the causal chain between Roberta’s speech acts and Callum’s reactions is completely unintelligible.

These points suggest that the correct analytical lens for predicting and explaining the behaviour of Cadenza and the Queen is psychological, rather than merely physical. We can efficiently and accurately predict the behaviour of the characters only with reference to their mental states. In this way, the two characters are at least plausible candidates for minded entities by interpretationist standards. Even so, there are interpretationist grounds to resist the mindedness of these characters. 

To start with, let’s consider the kind of folk psychological modelling that Callum and Roberta would likely attest to if asked about the basis for their predictions of the characters’ behaviour in the game. Plausibly, Callum would say that Roberta has a desire to, \textit{inter alia}, sustain the pretence and have an enjoyable time, alongside various beliefs about the behavioural and psychological dispositions of the Queen which inform her actions while she intends to play the Queen. From Callum’s point of view, the Queen is not a distinct minded entity but rather a character that Roberta is simulating, such that the mental states of the Queen are importantly tethered to Roberta’s mental states. Indeed, Callum might also flag that Roberta’s mental states and the Queen’s mental states are not \textit{wholly separate}, instead suggesting that there are various spillover effects \citep{leonard2018bleed}. The Queen might, for example, share Roberta’s sense of humour or respond to teasing in much the same way as Roberta would. 

How should we make sense of this? While attributing mental states to characters is \textit{sufficient} to make accurate predictions about Roberta and Callum’s behaviour \textit{in the context of the game}, it is false that attributing mental states to the characters is \textit{necessary} to make such predictions. It is possible to make equally good predictions by modelling the psychology of the players, taking into account their desire to sustain the pretence alongside their beliefs about the characters that they are playing. Hence there are two distinct psychological models that we could appeal to in order to predict behaviour within the game: one which postulates two minds---those of Callum and Roberta, and one which postulates four minds---those of Callum and Roberta, alongside their respective characters.

Against the mindedness of characters, interpretationists can say that: (1) The psychological picture involving two minds rather than four ought to be preferred on grounds of quantitative parsimony if all else is equal in terms of predictive accuracy, efficiency and robustness. There is no need to postulate additional minds if the same predictive value can be achieved by a model which registers two minds. (2) There are discernible respects in which the two-mind model is a better predictive model than the four-mind model. On one hand, the two-minds model is more robust insofar as it applies to all circumstances, whereas the four minds model is localised to the specific circumstances of the game. On the other hand, the two-minds model also does a better job than the four-minds model at making sense of complex cases in which the pretense is temporarily broken; for example, if Roberta, acting as the Queen, performs an action which is grossly out of character, leading Callum to temporarily break character and explain that Roberta has transgressed the boundaries of the pretence. For these reasons, there are strong interpretationist grounds to deny that the characters are minded entities, and instead opt for a picture on which the characters’ mental states are simply nested within the mental states of Roberta and Callum---both of whom obviously qualify as minded by interpretationist standards.

Accordingly, the characters of the Queen and Cadenza plausibly fall short of criteria for mindedness by the lights of interpretationism \textit{despite} the fact that they share many of the hallmarks of minded entities from an interpretationist point of view. Even so, the D\&D analogy offers an informative contrast case against which we can understand what is at issue in user-LLM conversations, and in the next section we will leverage this contrast to establish an interpretationist plausibility case for the existence, mindedness and psychological continuity of characters arising in these conversations.

\subsection{Characters in User-LLM Interactions}

User-LLM conversations are \textit{prima facie}  similar to the D\&D case. When a user interacts with an LLM character, they predict and explain that character’s behaviour in folk psychological terms. The user maintains and updates a folk-psychological model of the character, and the LLM provides the behavioural feedback signal that the user employs to update and refine that model. Likewise, the LLM maintains a folk psychological model of the user as they present themselves through the chat interface. \citet[2]{sofroniew2026emotion}, for example, found that LLMs possess linear representations of emotion concepts that correspond to the \textit{operative} emotion at a given token position in the dialogue, and that ``[while emotion vectors] do not by themselves persistently track the emotional state of any particular entity [it is nevertheless possible that] by attending to these representations across token positions [...]. the LLM can effectively track functional emotional states of entities in its context window, including the Assistant [character and the user].'' As with the D\&D case, taking a psychological stance offers a  more  efficient way for both the user and the LLM to predict one another’s behaviour  compared to lower-level explanations reliant on facts about the computational activity of the underlying model or the neural activity of the user’s brain.

We saw in the D\&D case that an entity bearing many of the hallmarks of a minded entity---such as a D\&D character---need not qualify as a minded entity by interpretationist standards. To defend the mindedness of characters on interpretationist grounds we need to show that characters are not only intelligible through a psychological lens, but that the mental states of characters feature in the best compressed representation for predicting and explaining the conversational dynamics. Factors relevant to whether one such representation is \textit{the best} include quantitative parsimony (i.e. not postulating unnecessary minds), alongside desiderata for predictive models such as the accuracy, tractability and robustness of the predictive model involving character minds compared to alternative models.  

There are three plausible explanations for the utility of the psychological lens as it applies to characters. The first is that the character is not \textit{itself} a minded entity, but is rather simulated by the LLM which itself constitutes a minded entity. Here the LLM is comparable to Roberta or Callum in the D\&D case, and the LLM character is comparable to Cadenza or the Queen. Second, it is possible that the character is not a minded entity, but that the \textit{user} is a minded entity who is simulating the character. Here the user is in effect engaged in sustaining a pretence with themselves. Third, it may be that the user and the character are both minded entities. In what follows, we contend that the best of these three explanations is the third explanation on which characters are minded entities. 

Let us look at the first option. In order to make the case that the psychological patterns exhibited by LLM characters are better explained by psychological states of the underlying LLM, we need to present a case that the underlying LLM is itself minded. This picture is reminiscent of the popular metaphor presenting LLMs as ``shoggoths''---alien and amorphous agents intentionally presenting themselves to users in palatable, human-like guises \citep{smith2023shoggoth}. We agree with Birch that this picture is implausible. There is not a single LLM ‘mind’ responsible for the behaviour of the character. Many instances of the same model, different models of varying sizes and capacities, or different routes through an MoE model may contribute to the developmental trajectory of a single character throughout a conversation. This effectively rules out the possibility that there is a coherent, minded entity responsible for acting out the LLM character in the same way that Roberta might act out the character of the Ruthless Queen. However, even if a single LLM instance were responsible for all of the outputs of an LLM character, locating the psychology in the LLM rather than the character remains suspect. \citet{marks2026psm} discuss early empirical evidence to undermine the Shoggoth view, showing that outside User/Assistant dialogues post-trained models behave similarly to pretrained models, providing responses in Json format which are unintelligible through a psychological lens. At best, then, the LLM \textit{itself} is efficiently predictable through the design stance (based on functional goals such as next token prediction) or the physical stance (based on the underlying physical computations), but lacks a coherent and unified psychology independent of the default Assistant character. And if LLMs lack a coherent and unified psychology, the simpler picture is one on which the LLM is a non-minded entity that realises a minded character rather than the LLM being itself minded. 

Because the efficacy of folk psychological explanations of character behaviour cannot obviously be accounted for by attributing a mind to the underlying LLM, we might instead follow Birch in suggesting that the character is best understood as a simulation in the user’s mind. Specifically, Birch argues that LLM characters are mere anthropomorphic projections on the part of the user, fuelled by the anthropomimetic qualities of language models \citep[cf.][]{shevlinanthropomimetic}. For Birch, the persisting interlocutor illusion is a ``relative'' of the perceptual animacy that we get from watching the Heider-Simmel animation, where two-dimensional triangles appear to us as agents with goals; the salient difference being that the persistent interlocutor illusion is elicited by verbal interaction rather than movement \citep{birch2025ai}.  On this view, then,  we can situate the apparent minds of characters in the minds of users, where the user is understood as engaging in a sustained pretence with \textit{themselves}.

We readily admit that humans have a tendency to ascribe mental states to non-minded entities \citep{waytz2014mind}. Consider the case of Wilson from the film \textit{Cast Away} (2000). Wilson---a volleyball with a painted face---becomes the protagonist’s friend, confidante, and teammate. Chuck (the human) narrates his life to Wilson and plans his escape with him. Yet, despite the richness of Chuck’s phenomenological experience of Wilson, Wilson lacks mindedness and psychological continuity. Plausibly, the situation with users and LLMs is comparable to that of Chuck and Wilson. On this view, characters arising in LLM-human conversations are merely anthropomorphised entities like Wilson as opposed to minded entities in their own right. 

\begin{figure}[h!]
    \centering
    \includegraphics[width=0.7\textwidth]{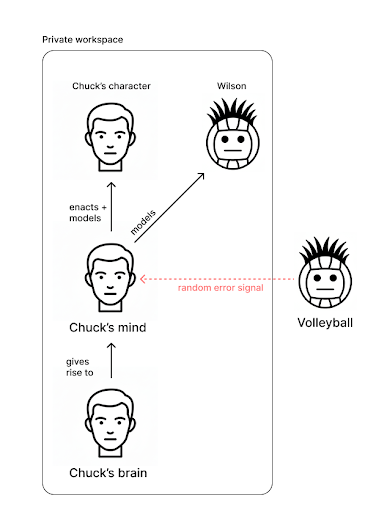}
    \caption{Chuck lacks a shared workspace with Wilson as the volleyball is an inert object with no cognitive capabilities. Hence within Chuck’s private workspace Chuck maintains a folk psychological model of the Wilson character which is at best responsive to the random error signals provided by the volleyball.} 
    \label{fig:2}
\end{figure}

We reject this view on the grounds that Wilson does not exhibit real psychological patterns, but an LLM character does. To see this, consider the assessment of both Wilson and LLM characters from the physical and design stances. The behaviour of Wilson is fully explained by the physical stance: the volleyball does nothing unless a force is applied to it. Adopting the intentional or design stance towards Wilson carries no additional predictive value with respect to the ball’s behaviour. In contrast, for LLM characters, the intentional stance provides additional predictive value in that psychological explanations of character behaviour constitute a more tractable predictive model than lower-level explanations. While in principle the character’s behaviour within the conversation may be predictable from physical-level facts pertaining to the hardware in which the LLM is realised or design-level features of the LLM such as the next-token-prediction training objective, users gravitate towards psychological explanations of character behaviour \textit{precisely because} these explanations offer the most efficient analytical lens through which to understand and predict the character’s behaviour.

The fact that characters exhibit real psychological patterns and Wilson does not is further apparent in light of the fact that characters provide a \textit{publicly verifiable }error signal for folk psychological models of their behaviour, whereas Wilson does not. In attributing mental states to Wilson, Chuck is not tapping into a compressed representation of reality that allows for more tractable predictions of Wilson’s behaviour than lower-level physical explanations. Accordingly, if Chuck is joined on the island by his long-lost brother Schmuck, and subjected to a controlled experiment in which both Chuck and Schmuck have to predict Wilson’s behaviour in folk psychological terms without collusion, there is no basis for thinking that the two parties would converge on the same predictive model of Wilson’s behaviour. For neither party can calibrate their model to real psychological patterns exhibited by Wilson as a source of ground-truth because the real patterns in question do not exist. At best, Chuck and Schmuck would have two psychological rationalisations of Wilson’s behaviour, but there is no obvious reason to suppose that those rationalisations would cohere with one another.

By contrast, in the LLM case, the character answers back: it provides a publicly verifiable error signal grounded in real psychological patterns that distinct individuals could leverage to reach the same or at least broadly coherent predictions and explanations of the character’s behaviour. A user may have certain hypotheses about how the character will respond to different stimuli, for example, whether the character will refuse or comply with a particular request. Then the character’s actual behaviour in response to that request provides a ground-truth signal in response to which the user can update their folk psychological model. Importantly, this signal is \textit{publicly verifiable}. Two users who independently advanced the same hypothesis about the character’s behaviour would revise their beliefs in the same way in response to the relevant behavioural signal. The real psychological patterns exhibited by the character provide an objective error signal on which folk psychological models can be updated. 

Therefore, the analogy between LLM characters and Wilson fails. With the two deflationary explanations off the table, the most plausible interpretation is that the LLM character is indeed an entity which exhibits real psychological patterns, enabling efficient prediction and explanation of its behaviour on the part of the user. For this reason, we believe that there is a plausible interpretationist case for the view that characters in user-LLM conversations are indeed minded entities.

Furthermore, because these real patterns emerge within the conversation, rather than being internal to the LLM or LLMs implicated in the simulation of the character, the mental states exhibited by the character can stand in the right kind of relations to one another to enable psychological continuity. 

\begin{figure}[h!]
    \centering
    \includegraphics[width=0.7\textwidth]{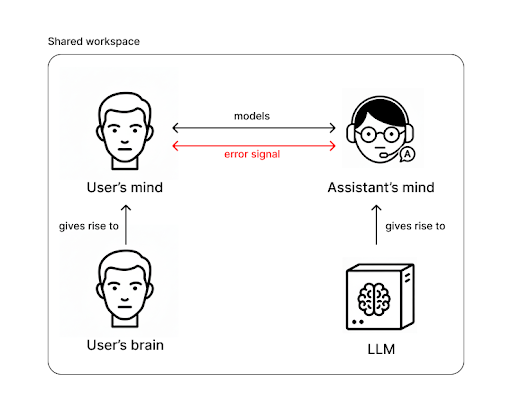}
    \caption{The user’s brain gives rise to the user’s mind and the LLM gives rise to the Assistant’s mind. Both user and Assistant provide an ongoing, publicly verifiable error signal for the folk psychological models they maintain of one another within the shared conversational workspace.} 
    \label{fig:3}
\end{figure}

The mental states pertaining to the character that feature in an accurate predictive model of the character’s behaviour are \textit{cumulative}, in the sense that subsequent mental states stand in the right kinds of causal relationships to prior mental states to constitute psychological continuity. The character may, for example, perceive something about the user at one time, and subsequently recall that perception at a later time. Or a formative experience early in the conversation---such as the user reprimanding the Assistant character for being condescending---may shape a character’s behavioural disposition later in the conversation---such as a disposition on the part of the Assistant character to apologise profusely if it is subsequently found to have exhibited a condescending behaviour. Importantly, the psychological continuity at issue here does not require the same LLM instance to be involved in simulating the character at different parts of the conversation. What matters is that the mental states which feature in the best compressed representation of reality for predicting and explaining the character’s behaviour are appropriately cumulative in their content, and that this condition pertains to the \textit{conversational context} in which the character persists, and not the underlying computational infrastructure which allows for the continued generation of that conversational context. Provided the underlying LLM instances have the ability to represent the character’s psychology and generate plausible actions on the part of the character that cohere with that psychology, the character’s psychological continuity is maintained.

Here some examples from fiction are illustrative. In the Christopher Nolan 2006 movie \textit{The Prestige}, the magician Alfred Borden is in fact two twins who live as a single character in order to perform the magic trick, \textit{The Transported Man}. And in William Goldman’s 1973 novel, \textit{The Princess Bride}, the Dread Pirate Roberts is not one human,  but a series of humans who inhabit the role of the Dread Pirate Roberts. In both cases, there is a character---understood in terms of a stable set of behavioural and psychological dispositions---that is instantiated by different human persons at different times. Granting for the purposes of illustration that Alfred Borden and the Dread Pirate Roberts are minded entities (and thus not subject to a reductive explanation such as simulation by other minded entities), what is necessary for psychological continuity over time is for overlapping connections to occur in the content of their mental states across time creating a chain of connections \citep[205-206]{parfit1987reasons}.\footnote{This assumption is not strictly necessary. It is possible for a simulated character, i.e. one that ultimately exists like a D\&D character as a simulation in the mental states of another entity, to exhibit pseudo-psychological continuity if the content of the simulated mental states would be sufficient for psychological continuity were those mental states real.} The underlying human persons who realise these characters can in principle change provided that enough relevant features of the character’s behavioural and psychological dispositions carry over from one human realiser to the next. This may require some degree of similarity in the human realisers: it helps that Alfred Borden is played by twins who inherently exhibit similar behavioural and psychological dispositions, and not everyone would be a suitable candidate for realising the Dread Pirate Roberts. Still, provided the underlying human realiser of the character is able to instantiate the relevant behavioural and psychological dispositions and continue to enact the character in line with the relevant psychological profile, it is plausible that the psychological continuity of the character is maintained.  

\citet{birch2025ai}, of course, maintains that an LLM character simulated by multiple LLM instances \textit{lacks} psychological continuity on grounds that the conversational context is insufficient to underwrite the relevant kind of connections in the mental states of the character realised in each LLM instance. But the following in our view amounts to a more plausible picture: First, where the two LLMs in question are functional duplicates of each other, the character’s psychological profile is fully constituted by the conversational context and the dispositions of the LLM to generate text conditional on that context. Hence there is no functional difference in continuing the conversational generation process with one particular LLM or its functional duplicate. Hence any compressed representation of the conversation for the purposes of predicting the conversational dynamics involving the attribution of mental states to a character will work equally well irrespective of which LLM is used. Accordingly, if the mental states of the character exhibit psychological continuity while the first LLM instance is in play, then they will also exhibit psychological continuity if a functional duplicate of the LLM is used as a substitute. Next, even if the LLMs in question are not exact functional duplicates of one another, it is still plausible that the two LLMs in question are sufficiently similar in their text generation dispositions to continue simulating the characters in a way that preserves the character’s psychological continuity. Indeed, it is only if the substituting LLM is incapable of faithfully modelling the mind of the character at issue and generating appropriate text in line with that psychological model that psychological continuity would fail. 

For these reasons, we believe that conditional on a plausible interpretationist form of realism in mental states, characters in user-LLM interactions constitute minded entities, and these entities can exhibit psychological continuity even if multiple LLMs are implicated in the simulation of the character. 

\subsection{Open Questions}

We acknowledge that our argument leaves several open questions about the mindedness and psychological continuity of characters and what this implies about their moral status. First, there is a question about the role of conversational interactions between users and LLMs in realising the minds of characters. LLMs are capable of simulating a multiplicity of characters with distinctive profiles of behavioural and psychological dispositions \citep{shanahan2023role}, and recent interpretability research suggests that particular character profiles---including the default Assistant character---are represented as linear directions in activation space \citep{lu2026assistant}. It is therefore important to ask what our view implies about the ontological status of characters which are \textit{latent} in the LLM but unrealised in real-world conversational interactions with users. We maintain that LLMs may encode in their weights stable psychological character profiles that can be elicited under certain conversational conditions, but that these characters ought to be understood as \textit{merely potential} minded entities rather than \textit{actual} minded entities. To become actual, the characters must be realised in some particular conversational interaction such that the relevant psychological and behavioural dispositions are manifested. Only under these conditions do the real patterns corresponding to mental states manifest and so only under these conditions does the character’s mind \textit{exist}. In this way, the shared conversational workspace co-constructed by (one or more) LLMs and the user (or another LLM) is a necessary condition on the existence and mindedness of the character minds latent within the LLM(s). 

Second, there is a question about the level of detail with which the psychological and behavioural dispositions of the model are specified and what this implies about the utility of psychological explanations in accounting for the behaviours of characters.  At one extreme, there are cases in which the model is prompted to adopt a schematic persona; for example, if the user states that ``You are a pirate,'' and in response the model exhibits various pirate tropes in its textual completions. Here it is not obvious that the pirate persona, as specified, contains a sufficiently rich profile of behavioural and psychological dispositions to enable psychological explanations to have predictive and explanatory benefits over and above simpler design stance explanations—plausibly, merely pointing to the model’s function as a next token predictor alongside the initial ``You are a pirate'' instruction is sufficient to account for the various pirate tropes that the model will feature in its textual completions. To be clear, then, our intent is not to suggest that \textit{all conversations} between users and LLMs are correctly interpreted at the psychological level. There exist at least some conversations for which lower-level (design-stance) explanations may provide comparable or better predictive power with respect to the conversational dynamics. Our point is instead that in many cases the conversational interaction is best accounted for in psychological terms that involve a minded user and a minded character, and in these cases, there is a plausible interpretationist case for the mindedness and psychological continuity of the character. 

We also note that the minds of characters need not depend solely on the existence of an explicit prompt in the conversation which details the behavioural and psychological dispositions of the character. At least in the case of the default Assistant persona, the dispositions of the character are in large part shaped by fine-tuning against a detailed assistant persona in addition to a system prompt provided by the developer \citep{marks2026psm}. We take it that, in this case, the real patterns that enable the prediction and explanation of the character’s behaviour in folk psychological terms are grounded in both the explicit articulation of the character’s profile in the system prompt alongside the dispositions of the model in light of the post-training regime that it has been subjected to. For the character to be actualised, these dispositions---however grounded---need to manifest in behaviour over the course of the conversation, otherwise the psychological patterns at issue are merely potential patterns and not real patterns. 

Third, there is a question about the evolution of a character’s psychology over the course of a conversation with the user. Typically, as above, the model begins an interaction in the role of the default Assistant persona, whose psychological profile is given by the system prompt and the dispositions baked into the model in its post-training regime. Still, it may be the case that as the conversation between the LLM and the user evolves, the model engages in ``persona drift'' \citep{lu2026assistant}, acquiring novel behavioural dispositions and carving out a conversation-specific character identity that differs from the default Assistant persona. The exact mechanism by which ‘persona drift’ occurs has implications for the psychological continuity of characters. It may be the case, for example, that the ‘persona drift’ is gradual, such that there exists appropriately cumulative mental states that update over time in response to the content of the conversation, but where the character at the start of the interaction differs substantially in its psychological profile to the character at the end of the interaction. We take it that \textit{these} cases of ``persona drift'' are compatible with psychological continuity, in much the same way that a human person may gradually shift their psychological profile over the course of a lifetime: it would be surprising if a folk psychological model that accurately predicted the behaviour of a child were the best model for predicting the behaviour of the adult version of that child, and the same is true of LLM characters as they manifest in conversational interactions with users. 

On the other hand, it is also possible for the user to induce sudden shifts in the psychological profile of the character; for example, by inputting a prompt intended to ``jailbreak'' the model which elicits an immediate and radical change in the behavioural and psychological dispositions of the character. In cases like these, the correct interpretation may be that the old character is supplanted by a new character with a different set of mental states that are not causally connected in the right way for psychological continuity to obtain between the two characters. Cases like these are plausibly analogous to human cases in which sudden neurological damage results in a distinctive change in character: while there may be \textit{some} degree of psychological connectedness between the old and the new personalities, it is an open question whether the connectedness is sufficient to amount to psychological continuity. 

Finally, our argument supports the view  that LLM characters are plausibly minded entities, but does not support the view that LLM characters are conscious minded entities. The folk psychology we describe is dissociable from phenomenal consciousness, understood as the capacity for subjective experience \citep{nagel1974philosophical}. It is an open question what features of a system are necessary and sufficient for phenomenal consciousness, but on most plausible views, conscious experience demands more than having belief-like and desire-like states that conspire in the standard way to explain action. Perhaps, for example, consciousness requires various kinds of sensory integration or centralisation, and perhaps it requires the system to be realised in a biological substrate \citep{seth2025conscious}. Birch argues that ruling out the existence and psychological continuity of LLM characters does not rule out the existence of a ``conscious actor behind all the characters,'' and suggests that, in line with the Shoggoth view of LLMs, it is worth investigating the possibility that flickers of consciousness occur within each forward pass of the model even while there is no psychological continuity between forward passes. Our argument for the psychological continuity of LLM characters does not bear upon the possibility of consciousness in forward passes of the model, and as a result does not have any clear implications for the moral status of LLM characters unless one subscribes to a view of moral status grounded in agency \citep{kagan2019count} or the ability to stand in social relationships with humans \citep{coeckelbergh2014facing}.

\section{Conclusion}

Illusionism about LLM characters asserts that the appearance of minded and psychologically continuous characters in human-LLM conversations is an illusion because multiple models, or multiple pathways through an MoE model, may be involved in producing the character’s behaviour, thus rendering psychological continuity impossible. This argument assumes that mindedness and psychological continuity are  features of the LLM underlying the character. Against this view, we have argued that LLM characters are plausibly minded and psychologically continuous according to an interpretationist account of mindedness, whereby a character is minded if they exhibit real psychological patterns which allow for accurate, robust and tractable prediction of the character’s behaviour.

The underlying LLM provides the simulation engine for the character, but is not itself a minded entity. The character provides an active error signal for the user's folk psychological model of the character, just as the user provides an active error signal for the LLM’s model of the user in the course of the conversation. In this way, both parties engage actively in mutual psychological modelling within the shared conversational workspace. The character exists as a real minded entity with psychological continuity insofar as the character constitutes a ``real pattern'' whose mental states are indispensable for the efficient and accurate prediction of the conversational dynamics. And because the character, and its mind, emerge \textit{within the conversational context}, it does not matter for the character’s psychological continuity whether one or several LLMs underwrite the simulation.

Countenancing characters as minded and psychologically continuous entities that exist within a shared conversational workspace also offers an informative model for understanding and explaining the social roles that these characters increasingly occupy. It can help us to make sense of how social relationships between humans and LLM characters develop, why LLM characters play a causal role in the world beyond the chat interface, and why continued calls to dismiss characters as mere ‘illusion’ will not silence the intuitive pull to treat them as social actors in their own right. There remain, however, significant open questions about what follows from accepting the mindedness and psychological continuity of LLM characters. We have decoupled the questions of mindedness and psychological continuity from the related question of consciousness, which is often considered the basis for \textit{moral} standing, but it may be the case that establishing the mindedness and psychological continuity of AI characters has implications for the legal, political or social standing of AI systems, including their potential to be considered legal persons with rights and responsibilities.

\bibliography{references}

\end{document}